\documentclass{appolb}%
\usepackage{epsfig}
\usepackage{graphicx}
\usepackage{dcolumn}
\usepackage{bm}
\usepackage{amssymb}
\usepackage{latexsym}
\usepackage{amsmath}
\usepackage{amsfonts}%
\begin{document}

\title{Influence of Vector Mesons on the $f_0(600)$ Decay Width in a Linear Sigma Model
with Global Chiral Invariance\thanks{Presented by D. Parganlija at Excited QCD, 8-14 February
2009, in Zakopane (Poland)}}
\author{Denis Parganlija$^{a}$, Francesco Giacosa$^{a}$ and Dirk H. Rischke$^{a,b}$
\address{$^a$Institute for Theoretical Physics, Goethe University,
Max-von-Laue-Str.\ 1, D--60438 Frankfurt am Main, Germany \and $^b$Frankfurt Institute for Advanced Studies, Goethe University,
Ruth-Moufang-Str.\ 1, D--60438 Frankfurt am Main, Germany} }
\maketitle

\begin{abstract}
We consider a globally invariant chiral Lagrangian that contains vector and axial-vector mesons. We compute the $f_0(600) \rightarrow \pi \pi$ decay width and $\pi \pi$ scattering lengths and compare with the corresponding results in which the (axial-)vector degrees of freedom decouple. We show that the role of vector mesons has a great impact on these quantities.

\end{abstract}



\section{Introduction}

One way to investigate Quantum Chromodynamics (QCD) at low energies is by using effective models - such as the Linear Sigma Model - that possess the same global symmetries as QCD, most notably the chiral $SU(N_{f})_{r}\times SU(N_{f})_{l}$ symmetry (with $N_{f}$ flavours). In these effective models hadronic degrees of freedom, rather than quark and gluon degrees of freedom, are present; spontaneous breaking of the chiral invariance leads to the emergence of nearly massless pseudoscalar Goldstone bosons (e.g., pions for $N_{f}=2$). Their chiral partners, the scalar states, remain massless.

If the Linear Sigma Model is used, there are at least two possibilities to
choose the hadronic degrees of freedom in the corresponding Lagrangian: with
(pseudo-)scalar degrees of freedom only, or with the addition of
(axial-) vector degrees of freedom. In this paper we present a linear sigma
model \emph{with} vector and axial-vector mesons for $N_{f}=2$ and discuss two
important quantities: the $\pi\pi$ scattering lengths and the decay width of
$f_{0}(600)$ (in our notation: $\sigma$) into two pions. In particular, we aim
to investigate the role of (axial-)vector mesons for these quantities. To this 
end we investigate the limit in which the (axial-)vector mesons decouple and 
compare the results with the results of the full model.

The paper is organized as follows: in Section 2 the model is introduced, in Section 3 the results with and without vector mesons are presented and in Section 4 we present our conclusions.

\section{The Model}

The model is described by a $SU(2)_{r}\times SU(2)_{l}$ chirally symmetric
Lagrangian \cite{Reference1, Boguta, Mainz}:
\begin{align}
\lefteqn{\mathcal{L}=\mathrm{Tr}[(D^{\mu}\Phi)^{\dagger}(D^{\mu}\Phi
)]-m_{0}^{2}\mathrm{Tr}(\Phi^{\dagger}\Phi)-\lambda_{1}[\mathrm{Tr}%
(\Phi^{\dagger}\Phi)]^{2}-\lambda_{2}\mathrm{Tr}(\Phi^{\dagger}\Phi)^{2}%
}\nonumber\\
&  -\frac{1}{4}\mathrm{Tr}[(L^{\mu\nu})^{2}+(R^{\mu\nu})^{2}]+\frac{m_{1}^{2}%
}{2}\mathrm{Tr}[(L^{\mu})^{2}+(R^{\mu})^{2}]+\mathrm{Tr}[H(\Phi+\Phi^{\dagger
})]\nonumber\\
&  +c(\det\Phi+\det\Phi^{\dagger})-2ig_{2}(\mathrm{Tr}\{L_{\mu\nu}[L^{\mu
},L^{\nu}]\}+\mathrm{Tr}\{R_{\mu\nu}[R^{\mu},R^{\nu}]\})\nonumber\\
&  +\frac{h_{1}}{2}\mathrm{Tr}(\Phi^{\dagger}\Phi)\mathrm{Tr}[(L^{\mu}%
)^{2}+(R^{\mu})^{2}]+h_{2}\mathrm{Tr}[\vert L^{\mu}\Phi
\vert ^{2} + \vert \Phi R^{\mu} \vert ^{2}]\nonumber\\
&  +2h_{3}\mathrm{Tr}(\Phi R_{\mu}\Phi^{\dagger}L^{\mu})+...\text{ ,}
\label{Lagrangian}%
\end{align}
where $\Phi=(\sigma+i\eta_{N})\,t^{0}+(\vec{a}_{0}+i\vec{\pi})\cdot\vec{t}$
(scalar and pseudoscalar degrees of freedom), $L^{\mu}=(\omega^{\mu}%
+f_{1}^{\mu})\,t^{0}+(\vec{\rho}^{\mu}+\vec{a}_{1}^{\mu})\cdot\vec{t}$ and
$R^{\mu}=(\omega^{\mu} - f_{1}^{\mu})\,t^{0}+(\vec{\rho}^{\mu} - \vec{a}_{1}^{\mu
})\cdot\vec{t}$ (vector and axial-vector degrees of freedom; our model is
currently constructed for $N_{f}=2$ - thus, our eta meson $\eta_{N}$ contains
only non-strange degrees of freedom); $t^{0}$, $\vec{t}$ are the generators of
$U(2)$; $D^{\mu}\Phi=\partial^{\mu}\Phi-ig_{1}(L^{\mu} \Phi -\Phi R^{\mu})-ieA^{\mu}[t^{3},\Phi]$ ($A^{\mu}$ denotes the photon field), $L^{\mu\nu
}=\partial^{\mu}L^{\nu}-ieA^{\mu}[t^{3},L^{\nu}]-(\partial^{\nu}L^{\mu
}-ieA^{\nu}[t^{3},L^{\mu}])$, $R^{\mu\nu}=\partial^{\mu}R^{\nu}-ieA^{\mu
}[t^{3},R^{\nu}]-(\partial^{\nu}R^{\mu}-ieA^{\nu}[t^{3},R^{\mu}]).$ The dots
refer to further globally invariant terms which are irrelevant in the following.
The explicit breaking of the global symmetry is described by the term
Tr$[H(\Phi+\Phi^{\dagger})]\equiv h\sigma$ $(h=const.)$. The chiral anomaly is
described by the term $c\,(\det\Phi+\det\Phi^{\dagger})$ \cite{Hooft}. Note
that the model possesses global chiral invariance; the reasons to use the
global invariance (instead of the local one) may be found, e.g., in Ref.
\cite{UBW} as well as Refs. \cite{Reference1, Parganlija:2008xy}. The reason to consider operators up to the fourth order only has been presented in Ref. \cite{dynrec}.

When $m_{0}^{2}<0$, spontaneous symmetry breaking $SU(2)_{r}\times
SU(2)_{l}\rightarrow SU(2)_{V}$ takes place. The sigma field is shifted by its
(constant) vacuum expectation value $\phi$: $\sigma\rightarrow\sigma+\phi.$ As a result of this shift non-diagonal mixing terms of the form $-g_{1}\phi \overrightarrow{a}_{1}^{\mu}\cdot\overrightarrow{\pi}$ and $-g_{1}\phi f_{1}^{\mu}\eta_{N}$ arise. In order to remove them, the fields $f_{1}$ and $\overrightarrow{a}_{1}$ are
also shifted as $f_{1}^{\mu}\rightarrow f_{1}^{\mu}+w\partial^{\mu}\eta$,
$\overrightarrow{a}_{1}^{\mu}\rightarrow\overrightarrow{a}_{1}^{\mu}%
+w\partial^{\mu}\overrightarrow{\pi}$ with $w=g_{1}\phi/m_{a_{1}}^{2}$ and the
pseudoscalar fields are renormalized: $\overrightarrow{\pi}\rightarrow
Z\overrightarrow{\pi}$, $\eta_{N}\rightarrow Z\eta_{N}$ where $Z=\frac
{m_{a_{1}}}{(m_{a_{1}}^{2}-g_{1}^{2}\phi^{2})^{\frac{1}{2}}}$ \cite{Mainz, RS}.

The identification of mesons with particles listed by the PDG \cite{PDG} is straightforward in the pseudoscalar and (axial-)vector sectors: The fields $\overrightarrow{\pi}$ and $\eta_{N}$ correspond to the pion and the $SU(2)$ counterpart of the $\eta$ meson, $\eta_{N}\equiv\sqrt{1/2}(\overline {u}u+\overline{d}d)$ (with a mass $m_{\eta_{N}}$\ of about $700$ MeV, obtained by 'unmixing' the physical $\eta$ and $\eta^{\prime}$ mesons). The
fields $\omega^{\mu}$, $\overrightarrow{\rho}^{\mu}$ represent the
$\omega(782)$ and the $\rho(770)$ vector mesons; the fields $f_{1}^{\mu}$, $\overrightarrow{a_{1}}^{\mu}$ represent the $f_{1}(1285)$ and $a_{1}(1260)$ axial-vector mesons. Unfortunately, the identification of the $\sigma$ and $\overrightarrow{a}_{0}$ fields is controversial with the possibilities being the pairs $\{f_{0}(600), a_{0}(980)\}$ and $\{f_{0}(1370),a_{0}(1450)\}$. For
the purpose of this paper we chose the assignment $\{f_{0}(600),a_{0}(980)\}$,
which allows us to study the decay $f_{0}(600)\rightarrow\pi\pi$ and the
$\pi\pi$\ scattering lengths.

Eleven parameters are present in the Lagrangian: $\lambda_{1}$, $\lambda_{2}$,
$c$, $h_{0}$, $h_{1}$, $h_{2}$, $h_{3}$, $m_{0}^{2}$, $g_{1}$, $g_{2}$,
$m_{1}$. Using the masses $m_{\pi}$, $m_{\eta_{N}}$, $m_{a_{0}}$, $m_{\rho}$ and $m_{a_{1}}$, the pion decay constant $f_{\pi}$ (via the Eq. $\phi=Zf_{\pi}$) and the experimentally well-known decay widths $\Gamma_{\rho\rightarrow\pi\pi}=149.4\pm1.0$ MeV and $\Gamma_{f_{1}\rightarrow a_{0}\pi}=8.748\pm2.097$ MeV \cite{PDG}, one can eliminate eight parameters. All
quantities of interest can be then expressed in terms of three independent
combinations of parameters, which for convenience are chosen to be $m_{\sigma
}$, $Z$ and $h_{1}$. Explicit expressions and a detailed description of the
procedure can be found in Ref. \cite{PGR}.

For the present purpose, it is useful to write explicitly the equation for the
$\rho$ mass%
\begin{equation}
m_{\rho}^{2}=m_{1}^{2}+\frac{\phi^{2}}{2}[h_{1}+h_{2}(Z)+h_{3}(Z)]
\label{rhomass}%
\end{equation}
where the decomposition of $m_{\rho}^{2}$ in terms of two terms is evident:
the term $\frac{\phi^{2}}{2}[h_{1}+h_{2}(Z)+h_{3}(Z)]$ represents the
contribution of the chiral condensate (and vanishes in the limit
$\phi\rightarrow0$), while $m_{1}^{2}$ is independent of the chiral condensate
and can be related to the gluon condensate. Note that we made explicit that
$h_{2}\equiv h_{2}(Z)$ and $h_{3}\equiv h_{3}(Z)$ are functions of $Z.$ In the following we will consider $m^2_1$ as varying between 0 and $m_\rho^2$: in fact, a negative $m^2_1$ would imply that the vacuum is not stable in the limit $\phi\rightarrow 0$; $m_{1}^{2}>m_{\rho}^{2}$ would imply that the contribution of the chiral condensate is negative. This is counter-intuitive
and at odds with various microscopic approaches such as the NJL model.

\section{Results and discussions}

\textit{\textbf{Results with (axial-)vector mesons: }}The scattering lengths $a_{0}^{0}\equiv a_{0}^{0}(m_{\sigma},Z,h_{1})$ and $a_{0}^{2}\equiv a_{0}^{2}(m_{\sigma},Z,h_{1})$ and $\Gamma_{a_{1}\rightarrow\pi\gamma}\equiv \Gamma_{a_{1}\rightarrow\pi\gamma}[Z]$ are functions of $m_{\sigma}$, $Z$ and $h_{1}$ (explicit expressions can be found in Ref. \cite{PGR}). We use the experimental
values $a_{0}^{0}=0.218\pm0.020$ and $a_{0}^{2}=-0.0457\pm0.0125$ from the
NA48/2 collaboration \cite{Bloch} and $\Gamma_{a_{1}\rightarrow\pi\gamma
}=0.640\pm0.246$ MeV \cite{PDG}. A standard fit procedure allows us to
determine the parameters $m_{\sigma}$, $Z$, and $h_{1}$ and their errors:
$Z=1.67\pm0.20$, $h_{1}=-68\pm338$ and $m_{\sigma}=(332\pm456)$ MeV.
Obviously, the errors of $h_{1}$ and $m_{\sigma}$ are too large. A drastic improvement can be obtained via a physical consideration: as we discussed after Eq. (\ref{rhomass}) we impose that $m_{1}^{2}$ is varying between $0$ and $m_{\rho}^{2}$. As a consequence for Z=1.67, the parameter $h_1$ needs to lie in the interval $h_{1}\in\lbrack -32,-83]\Longleftrightarrow h_{1}=-68_{-15}^{+36}$ (the upper boundary of $h_{1}$ corresponds to $m_{1}=0$). The value of $m_{\sigma}$ can be constrained in a similar way using the scattering length
$a_{0}^{0}$: $m_{\sigma}\in\lbrack279,500]$ MeV $\Leftrightarrow m_{\sigma}=332_{-53}^{+168}$ MeV.

With these parameter values it is possible to calculate the $\sigma$ decay
width, $\Gamma_{\sigma\rightarrow\pi\pi}\equiv\Gamma_{\sigma\rightarrow\pi\pi
}(m_{\sigma},Z,h_{1}),$ which is shown in the following \textit{Fig. 1}.

\begin{figure}[h]
\centering
\includegraphics[scale = 0.43]{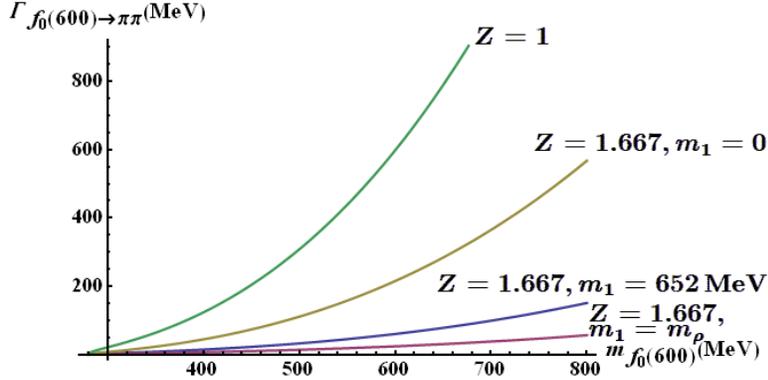}
\caption{Scattering length $a_{0}^{0}$ with ($Z=1.667$) and without ($Z=1$) vector and axial-vector mesons}
\end{figure}

The three lower curves in the plot depict the $\sigma$ decay width in the
model that contains vector and axial-vector degrees of freedom. It is clear
that the decay width becomes smaller when the vector and axial-vector degrees
of freedom are introduced. In that case, the value of the $\sigma$ decay width changes very slowly with $m_{\sigma}$ so that even for $m_{\sigma}\cong 650$ MeV the width is not larger than approximately 300 MeV - a value that is too small when compared, e.g., with the results from Refs. \cite{Leutwyler, Pelaez1}. Note that the value of $m_{\sigma}\cong 650$ MeV is actually too large to agree with the scattering length $a_{0}^{0}$ (as apparent from {\textit Fig. 2}) if the (axial-)vector fields are considered - then, the $\sigma$ mass should not be larger than 450 MeV which in turn yields a $\sigma$ decay width smaller than about 100 MeV.

\begin{figure}[h]
\centering
\includegraphics[scale = 0.37]{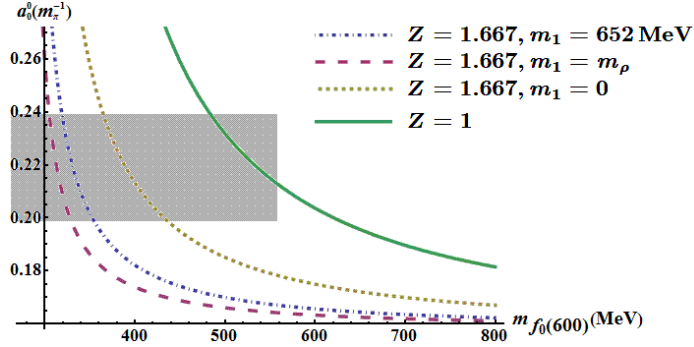}
\caption{Scattering length $a_{0}^{0}$ with ($Z=1.667$) and without ($Z=1$)
vector and axial-vector mesons. The shaded area corresponds to NA48/2 results \cite{Bloch}}%
\end{figure}

 $\,$\\

\textit{\textbf{Results without (axial-)vector mesons: }}This limit is
obtained for $g_{1}\rightarrow 0$, i.e. $Z\rightarrow 1$, and $h_{1}=h_{2}%
=h_{3}=0.$ In this case the decay width $\Gamma_{\sigma \rightarrow \pi \pi} \equiv \Gamma_{\sigma \rightarrow \pi \pi}(m_\sigma, Z = 1, h_1 = 0)$ and the scattering length $a_0^0 \equiv a_0^0(m_\sigma, Z = 1, h_1 = 0)$ are functions of $m_\sigma$ only and are plotted in {\it Fig. 1} and {\it Fig. 2}, respectively. The scattering length $a_{0}^{0}$ allows for $m_{\sigma}\in\lbrack500,620]$ MeV (see \textit{Fig. 2}). Thus, the value of the $f_{0}(600)$ decay width in this case is \emph{much larger} than in the previous cases: given that $m_{\sigma}\in\lbrack 500,620]$ MeV, one obtains $\Gamma_{\sigma}\in\lbrack 350,600]$ MeV (see {\it Fig. 1}). These values are in agreement with the result obtained by Leutwyler \textit{et al.} ($\Gamma_{\sigma}/2=272_{-12.5}^{+9}$ MeV) \cite{Leutwyler}
and the result $\Gamma_{\sigma}/2=(255\pm16)$ MeV obtained by Pel\'{a}ez
\textit{et al.} \cite{Pelaez1}. However, this limit also implies that
$\Gamma_{\rho\rightarrow\pi\pi}=\Gamma_{a_{1}\rightarrow\rho\pi}=0$, which is
clearly at odds with experiment \cite{PDG}.

\section{Conclusions}

The $\sigma$ model without (axial-)vector degrees of freedom can properly describe $\pi\pi$ scattering and $f_{0}(600)$ phenomenology. This is also one of the reasons for its success in low-energy hadron physics. However, the
inclusion of vector and axial-vector fields drastically worsens the agreement
with data: a simultaneous description of both $\pi\pi$ scattering lengths and
$f_{0}(600)\rightarrow\pi\pi$ is no longer possible. This is due to the fact
that a notable decrease in the decay width of the $f_{0}(600)$ resonance,
$\Gamma_{\sigma\rightarrow\pi\pi}\cong100$ MeV, takes place and due to the
important role of the $\rho$ meson in $\pi\pi$ scattering.

In light of these results one could be led to neglect (axial-)vector degrees
of freedoms. This is, however, not feasible: vector and axial-vector mesons
are well-known quark-antiquark states which should necessarily be part of a
realistic low-energy model. Neglecting them would imply that they decouple from the pseudoscalar Goldstone bosons, which is obviously at odds with the experimentally measured decay widths $\Gamma_{\rho\rightarrow\pi\pi}$ and $\Gamma_{a_{1}\rightarrow\rho\pi}$. A way out of the problem is to change
the scenario for scalar mesons by identifying the scalar fields of the model
with $\{f_{0}(1370),a_{0}(1450)\}$ (as done - for instance - in Refs.
\cite{Amsler:1995td,Fariborz:2005gm}). Still, a light scalar resonance with mass of about 400 MeV is needed to describe $\pi\pi$ scattering: the resonance $f_{0}(600)$ can be interpreted as an additional tetraquark state. In this way,
mixing of tetraquark and quarkonia takes place \cite{Fariborz:2005gm}. A detailed study of this possibility and the generalization to $N_{f}=3$ (involving also the scalar glueball) are possible extensions of the present approach.

\end{document}